\documentclass{ws-ijmpa}

\begin{document}

\markboth{RRW}
{Are there strangelets in cosmic rays?}

\catchline{}{}{}

\title{ARE THERE STRANGELETS IN COSMIC RAYS?}

\author{\footnotesize M. RYBCZY\'NSKI and Z. W\L ODARCZYK}

\address{Institute of Physics, \'Swi\c{e}tokrzyska Academy, ul.
\'Swi\c{e}tokrzyska 15,  25-406 Kielce, Poland}

\author{G. WILK}

\address{The Andrzej So\l tan Institute for Nuclear Studies,
Zd-P8, ul. Ho\.za 69; 00-681 Warsaw, Poland}

\maketitle

\pub{Received (Day Month Year)}{Revised (Day Month Year)}

\begin{abstract}
Assuming that cosmic rays entering the Earth's atmosphere contain a
small admixture of nuggets of strange quark matter in form of {\it
strangelets} one can explain a number of apparently "strange" effects
observed in different cosmic rays experiments. We shall demonstrate
here that the mass spectrum of such strangelets filles the "nuclear
desert" gap existing between the heaviest elements observed in
Universe and the next "nuclear-like objects" represented by neutron
and strange stars. 

\end{abstract}

\vspace{5mm}
Cosmic ray experiments of different kinds abound in results which are
rather unexpected when considered with the experience rooted in the
accelerator physics only \cite{Exotica1,Exotica2}. The most
spectacular observations are: anomalous cosmic ray bursts from {\it
Cygnus X-3}, extraordinary high luminosity gamma ray burst from the
{\it supernova remnant N49} in the Large Magellanic Cloud, or the so
called {\it Centauro} events, which are characterised by anomalous
composition of secondary particles with no neutral pions present. As
demonstrated in detail in \cite{Exotica2,S} substantial part of such
events can be explained by assuming that cosmic rays entering the
Earth's atmosphere contain a small admixture of nuggets of strange
quark matter (SQM) in form of {\it strangelets}, i.e., of nuggets
with $200 < A < 10^6$ (below lower limit stangelets become
unstable while above upper limit they contain also electrons and then
their size is limited only by the unstability due to gravitational
collaps which happens at $A\sim 10^{57}$). Such SQM is probably
continuosulsy produced in some neutron stars (or in quark stars). For
example, {\it Chandra observatory} has discovered an object, which
probably could be a strange star \cite{Stars}. One should mention
here that recently one event with small ratio $Z/A$ have been found
experimentally with AMS detector\cite{AMS} (albeit with $A$ very
small, estimated to be $A\simeq 17.5$, i.e., it would be a metastable
strangelet). 

\begin{figure}[h]
\centerline{\psfig{file=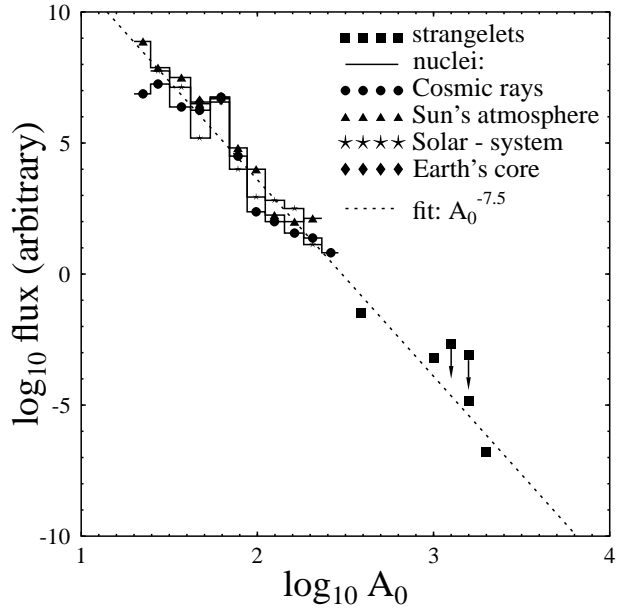,width=10cm}}
\vspace*{8pt}
\caption{Comparison of the estimated mass spectrum $N(A_0)$ of
strangelets with the known abundances of elements in the Universe
\protect\cite{Zhdanov}. Consecutive steps in histogram denote the
following nuclei (or groups of nuclei): Ne, (Mg,Si), S, (K,Ca),
Fe, (Cu,Zn), (Kr,Sr,Zr), (Te,Xe,Ba), (Rare earths), (Os,Ir,Pt,Pb).
The flux of strangelets used here has been obtained from attempt to
accomodate together all possible signals of strangelets
\protect\cite{Exotica2,S}.}  
\label{Fig1}
\end{figure}

In this presentation we would like to discuss one more
feature of strangelets, namely the fact that they seem to fill the
gap, or the {\it nuclear desert}, existing between the heaviest
elements observed in Universe and the next {\it nuclear-like objects}
represented by neutron stars (or possibly by strange and quark stars)
\cite{Stars,Z}. They do by following the empirical (not so well known)
$A^{-7.5}$ power law dependence found long time ago in \cite{Zhdanov}
when analysing the abundance of the elements in the Universe, see
Fig. \ref{Fig1}.    

So far this $A^{-7.5}$ dependence (and, in general, the reason for
such power law to show up) is, to our knowledge, unexplained. We
would like to propose here very simple reasoning leading to such
behaviour. Let us consider an object (nucleus or strangelet) carrying
mass number $A_i$, which absorbes neutrons with probability $W=\Phi
\sigma = A_i v\sigma$. In time $t$ it travels distance $L=vt$. If
absorption of neutrons proceeds with cross section $\sigma$ and if
density of neutrons is $n$, then the increment of mass of such object
in a unit time is $\partial A_i = nA_iL\sigma \partial t/t$. For the
$i$-th such object one has then that

\vspace{-14mm}

\begin{equation}
\frac{\partial A_i}{A_i}\, =\, \alpha\cdot \frac{\partial t}{t}\qquad
\Longrightarrow\qquad A_i(t)\, =\,
m\left(\frac{t}{t_i}\right)^{\alpha} , \label{eq:A(t)}
\end{equation}

\noindent
where $\alpha = nL\sigma$. The initial condition used here is that
the $i$-th object occurs in time $t_i$ with mass $A_i(t_i) = m$.
Probability of forming an object of mass $A_i(t) < A$ is then

\begin{equation}
P(A_i(t)<A)\, =\, p\left(t_i > \frac{m^{1/\alpha}\cdot
t}{A^{1/\alpha}}\right) . \label{eq:P(A)}
\end{equation}

\noindent
Assuming now uniform distribution of the occurence of such objects in
the system, i.e., that the probability of adding such object in the
unit time interval to the system is $P(t_i) = 1/t$, one gets  that

\begin{equation}
P\left(t_i > \frac{m^{1/\alpha}\cdot t}{A^{1/\alpha}}\right)\, =\,
1\, -\, P\left(t_i \le \frac{m^{1/\alpha}\cdot
t}{A^{1/\alpha}}\right)\, =\, 1\, -\,
\frac{m^{1/\alpha}}{A^{1/\alpha}} \, . \label{eq:ineq}
\end{equation}

\noindent
This, together with eq. (\ref{eq:P(A)}), results in the following
expression for the probability of forming an object of mass $A$:

\begin{equation}
P(A)\, =\, \frac{\partial P(A_i(t) < A)}{\partial A}\, =\,
\frac{m^{1/\alpha}}{1+1/\alpha}\cdot A^{-(1+1/\alpha)} . \label{eq:eqA}
\end{equation}

It means then that in this way we have obtained the following
time-independent power-like distribution of abundances of elements of
mass number $A$:

\begin{equation}
P(A) \propto\, A^{-\gamma}\qquad {\rm with}\qquad \gamma = 1 +
\frac{1}{\alpha} = 1 + \frac{1}{nL\sigma} . \label{eq:final}
\end{equation}

\noindent
In our case $\gamma = 7.5$ corresponds to $\alpha = nL\sigma = 0.154$
and this, for the typical value of the cross section entering here
equal to $\sigma = 30$ mb, gives the thickness of the layer in which
our objets are to be produced equal to $nL = 5.1\cdot 10^{24}$
cm$^{-2} = 8.5$ g/cm$^2$. Assuming now for the typical density of
neutrons value of $n \sim 10^{20}$ cm$^{-3} = 1.6\cdot 10^{-4}$
g/cm$^3$ and assuming that the so-called $r$-process nucleosynthesis
conditions are met \cite{M} one gets that this can be achieved with a
rather thin layer of thickness equal to $L=0.5$ km.

To summarize: we have reminded that SQM can still be interesting
topic of research, in particular in what concerns {\it strangelets}
and their properties. In particular we have provided a viable
explanation of the specific $A$-dependence of their abundance in
terms of the usual nucler processes.

\end{document}